\documentclass[conference]{IEEEtran}
\IEEEoverridecommandlockouts
\usepackage{microtype}
\usepackage{amsmath,amssymb,amsfonts}
\usepackage{algorithm}
\usepackage{algorithmic}
\usepackage{graphicx}
\usepackage{textcomp}
\usepackage{xcolor}
\usepackage{hyperref}
\usepackage{orcidlink}

\newcommand{\SVR}{\ensuremath{\mathit{SVR}}}

\begin{document}
\title{Adaptive AI-based Decentralized Resource Management in the Cloud-Edge Continuum
\thanks{Corresponding author: Lanpei Li --- This work has been partially supported by the CHARITY H2020 project under EC grant ID 101016509.}
}


\author{
    \IEEEauthorblockN{
        Lanpei Li $^{\ast,\dagger}$ \orcidlink{0009-0005-4370-1020}, 
        Jack Bell $^{\dagger}$ \orcidlink{0009-0002-7857-4752}, 
        Massimo Coppola $^{\ast}$ \orcidlink{0000-0002-7937-4157}, 
        Vincenzo Lomonaco $^{\dagger}$ \orcidlink{0000-0001-8308-6599}
    }
    \IEEEauthorblockA{
        ${\ast}$\textit{ISTI-CNR}, Pisa, Italy \\
        \{lanpei.li, massimo.coppola\}@isti.cnr.it
    }
    \IEEEauthorblockA{
        ${\dag}$\textit{Department of Computer Science, University of Pisa}, Pisa, Italy \\
        jack.bell@di.unipi.it, vincenzo.lomonaco@unipi.it
    }
}

\maketitle

\begin{abstract}
In the Cloud-Edge Continuum, dynamic infrastructure change and variable workloads complicate efficient resource management. Centralized methods can struggle to adapt, whilst purely decentralized policies lack global oversight. This paper proposes a hybrid framework using Graph Neural Network (GNN) embeddings and collaborative multi-agent reinforcement learning (MARL). Local agents handle neighbourhood-level decisions, and a global orchestrator coordinates system-wide.  This work contributes to decentralized application placement strategies with centralized oversight, GNN integration and collaborative MARL for efficient, adaptive and scalable resource management.

\end{abstract}


\section{Introduction}
Traditional Cloud computing struggles to meet the quality-of-service (QoS) demands of IoT due to the scale, mobility, and geographical distribution of devices, as well as the need for low latency and localized processing\cite{shi_edge_2016}. The Cloud-Edge Continuum represents an evolution beyond traditional Cloud computing to address challenges arising from the rapid expansion of IoT devices and the vast amounts of data they generate. Unlike the cloud-centric model, the Continuum integrates a wide range of distributed computing resources across multiple layers, including cloud data centers, fog nodes, edge devices, and IoT endpoints\cite{bittencourt_internet_2018}. This architecture aims to support diverse application requirements, including  distributed computing to enable parallel processing across multiple nodes, scalability to accommodate fluctuating workloads, and energy efficiency for sustainable operation in resource-constrained environments. However, the Continuum introduces several challenges, including the need for efficient resource discovery, monitoring, and management, as well as the coordination of application execution across heterogeneous resources and environments\cite{carlini_smartorc_2023}.

Resource management in the Cloud-Edge Continuum is crucial for optimizing performance and ensuring seamless service delivery across distributed environments. The goals of resource management in the Cloud-Edge Continuum focus on optimizing resource allocation, load balancing, resource provisioning, task scheduling, and maintaining QoS\cite{mijuskovic_resource_2021}.

This paper proposes a decentralized application placement strategy for resource management in the Cloud-Edge Continuum. The framework leverages GNN to encode the states of resources and application components, providing a powerful representation for dynamic decision-making. Leveraging a collaborative MARL approach, local agents focus on optimizing resource management at the neighborhood level, while a global orchestrator facilitates coordination and high-level policy alignment. This integration of GNN-based embeddings and MARL enables the framework to balance scalability, adaptability, and accuracy in resource management.

\section{Problem Settings}
\label{sec:ps}

To evaluate resource management policies within the Cloud-Edge Continuum, three distinct models and a specified objective function, defined in \ref{sec:system-model}, are typically used:
\begin{itemize}
    \item \textbf{Resource Model:} depicts available computational, storage and network links capturing resource heterogeneity and dynamic changes (e.g. device capabilities)
    \item \textbf{Application Models:} defines the structure and interactions within multi-service applications, including the topology of their components and communication paths. 
    
    \item \textbf{Cost Model:} Considering completion time (CT) (computation and communication costs), resource utilization (RU) and SLA Violation Rate (SVR).
    \item \textbf{Objective Function:} optimizing balance of cost model parameters, such as RU, SVR and CT.
\end{itemize}

The dynamic nature of the Cloud-Edge Continuum, characterized by high variability, heterogeneity, and unpredictability, necessitates the use of autonomic control loops such as MAPE (Monitoring, Analysis, Planning, Execution) for effective resource management\cite{maurer_revealing_2011}. Models like resource, application, and cost models are integrated into the MAPE loop to guide decision-making in environments where resource states, and application demands fluctuate frequently.

While MAPE offers an automated approach for managing the Cloud-Edge Continuum, addressing the dynamic nature of the Continuum remains challenging. One major issue is accuracy, as static approaches often struggle to account for real-time changes in resource availability, network conditions, or application demands, leading to suboptimal decisions. Optimization overhead is another challenge, as dynamic environments necessitate frequent updates and recalculations, increasing computational costs and decision-making delays\cite{urgaonkar_dynamic_2015}. Centralized systems face scalability issues, often becoming bottlenecks when managing large, heterogeneous resources across multiple providers. Additionally, there are practical trade-offs between system responsiveness and long-term optimization, making it difficult to achieve a balance that satisfies all objectives. A decentralized, local-based approach could address these challenges by leveraging localized insights to improve accuracy, distributing computational tasks to reduce overhead, and enhancing scalability through collaborative management across multiple resource providers. However, decentralization introduces its own challenges, such as the need for effective synchronization to achieve global objectives and system-wide efficiency, making it essential to carefully design  communication mechanisms between local orchestrators such as the use of a global coordinator.

\section{Related Work}
\label{sec:rw}
Optimization-based approaches directly solving the optimization problem through methods such as Mixed Integer Linear Programming  (MILP) \cite{kyriakidis_milp_2012} is a well-researched problem area that has theoretical guarantees to find the optimal solution, but is only suitable for small-scale resource management problems.  Heuristic-based algorithms instead pose the task as a multi-dimensional bin-packing problem\cite{zhao_method_2024} such as first-fit, best-fit, worst-fit, and round-robin placement \cite{nandwal_enhancing_2021} strategies which have long been used for resource allocation tasks. 

Deep Reinforcement Learning (DRL) has been extensively explored for dynamic resource allocation, task scheduling, and autoscaling in Cloud-Edge Continuum environments. While centralized control can leverage a single DRL agent with a global view, decentralized control requires multi-agent reinforcement learning (MARL), as demonstrated in \cite{naderializadeh_resource_2021}, which used deep independent Q-learning (IQL) for distributed decisions in wireless networks. However, IQL struggles with non-stationary environments caused by other agents’ actions, a challenge addressed in \cite{lee_spqr_2023} through shared reward functions to improve stability. To handle high-dimensional data effectively, embeddings created using GNNs (e.g., GCNs \cite{yan_automatic_2020} and GraphSAGE \cite{hamilton_inductive_2017}) allow DRL agents to process variable-size graphs, as shown in \cite{randall_grows_2022}. While GNNs excel in representing complex systems, they require costly rebuilding when nodes are added or removed. Our approach mitigates this by masking unused nodes in the DRL agent state space, eliminating the need for frequent retraining.

Research on dynamic resource management has explored adaptive strategies to address changing application demands, network conditions, and resource mobility. For example, \cite{son_latency-aware_2019} proposed a latency-aware algorithm that uses edge and cloud resources for latency-sensitive applications, while \cite{ghobaei-arani_autonomic_2016} applied the MAPE control loop for real-time resource provisioning in cloud environments. RL techniques, such as the DRL-based algorithm by \cite{chen_deep_2021}, have also been used to optimize resource allocation and reduce task delays more effectively than traditional methods. 
Decentralized approaches distribute decision-making across local agents with limited information, interacting with neighboring nodes and agents. In \cite{ji2024graph}, GNNs and MARL enable autonomous decision-making in the V2X setting, where agents learn efficiently from local observations and align decisions using a weighted global reward to maximize the global objective. However, decentralization lacks a consistent global view, potentially leading to suboptimal convergence. By leveraging GNNs, MARL pretraining at the local coordinator level, and a global reward, we aim to mitigate the downsides of full decentralization. Additionally, MARL enables our approach to adapt to dynamic system conditions and to make flexible decisions. 


\section{Generic model}
\label{sec:gm}
Our proposed model begins with a centralized use case, where GNNs are employed during the Analysis phase of the MAPE autonomic control loop. The GNN serves as a feature extractor, capturing the state of individual nodes while accounting for their logical inter-dependencies. These extracted features are then utilized by a centralized DRL agent, which has full observability of the entire system. 

To efficiently address the inherent dynamism of the Cloud-Edge Continuum, this framework is extended to a decentralized approach by decomposing the problem into localized components. This decentralized design leverages a MARL Model, where multiple agents operate independently at the neighborhood level. These agents use localized insights to make decisions and collaborate to optimize resource allocation and system performance across the Continuum. This transition enhances scalability, adaptability, and responsiveness to dynamic changes in infrastructure and workloads.

\subsection{System Model}
\label{sec:system-model}
In the Cloud-Edge Continuum, distributed resources differ in computational power, memory, and network capabilities. Managing resources in this environment involves allocating application components to the available resources to optimize performance while meeting QoS requirements.
\subsubsection{Application Model}
An application is modeled as an undirected graph $G_A = \langle C, E \rangle$:
\begin{itemize}
    \item $C=\{c_1, c_2, \dots, c_N\}$ contains application components (microservices). Each component \(c_i\) requires a certain number of CPU/GPU cores, memory, storage, and has a deadline. 
    \item $E$ specifies communication constraints, including latency and bandwidth needs between components.
\end{itemize}

\subsubsection{Resource Model}
The resource model is modeled as $G_R = \langle V, L \rangle$, where:
\begin{itemize}
    \item V = $\{v_1,\dots,v_M\}$ represents computing nodes (e.g. edge devices, cloud data centres), each with available CPU/GPU, RAM, storage, and an availability indicator.
    \item $L$ describes network links with latency and bandwidth capacities.
\end{itemize}

\subsubsection{Cost Model}
\label{sec:cost_model}
\begin{itemize}
    \item \textbf{Completion Time (CT)} of a component $c_i$ on a resource $v_k$ combines computation and communication times. Computation costs encompass the time required for processing tasks, while communication costs include the delay involved in transferring data, influenced by data size, bandwidth usage, and network conditions. Total application completion time is the sum of all components' times, the upper bound is defined as the sum of computation and communication time.
    
    

    \item \textbf{Resource Utilization (RU)} is the average resource utilization across all nodes and edges considering CPU, GPU, etc.


    \item \textbf{SLA Violation Rate (SVR)} measures the fraction of components missing their deadlines. 


\end{itemize}

\subsubsection{Objective Function}
Maximizes RU while minimizing CT and SVR:
\begin{equation}
\text{Objective: } \max \left( \alpha \cdot RU - \beta \cdot CT_{app} - \gamma \cdot \SVR \right)
\end{equation}
Here, $\alpha, \beta, \gamma$ are weighting factors that determine the importance of RU, CT, and SVR, respectively. Future work will incorporate economic factors (e.g. operational costs, energy usage) to compare centralized and decentralized policies.

\section{Applying the model}
\label{sec:am}
\subsection{Agents}
We employ a hierarchical design with \textbf{local} and \textbf{global} agents that collaborate to optimize application placement in the Cloud-Edge Continuum:

 \subsubsection{Local Agents}
\begin{itemize}
    \item \textbf{Partial Observability:} Each local agent manages a partially observable neighbourhood subgraph(see Sect.\ref{sec:observationspace}).
    \item \textbf{Pretrained Behavior:} Pretrained using local reward to optimize efficiency within their subgraph.
    \item \textbf{Local Reward Function:} 
    \begin{equation}
    R_{local_i} = \alpha \cdot RU_{local_i} + \beta \cdot \frac{1}{CT_{local_i}} + \gamma \cdot \frac{1}{\SVR_{local_i}}
    \end{equation}
    balances RU, CT, and SVR for a given local agent.
\end{itemize}
\subsubsection{Global Agent}

\begin{itemize}
\item \textbf{Aggregate Observability:} Receives abstracted state information from local agents, detailed in Sect.\ref{sec:observationspace}. 
    \item \textbf{Joint Training:} Trained with local agents to align local and global objectives.
    \item \textbf{Global Reward} Considers non-local and local metrics:
    \begin{equation}
    R_{global} = \lambda \cdot R_{non-local} + \sum_{i} \mu_i \cdot R_{local_i}
    \end{equation}
    Where $R_{non-local}$ is defined similarly to $R_{local}$ but for system-wide metrics and $\lambda, \mu_i$ weight the importance of non-local and local objectives.

\end{itemize}

\subsection{Observation Space}
\label{sec:observationspace}
\textbf{Local Agent Observation} \newline
Each local agent has partial observability: it only sees the subgraph of $G_R$ corresponding to its local neighbourhood plus the full application graph. More formally:
\begin{equation}
O_{local} = \{S_{app}, S_{res}^{local}\}
\end{equation}
where the application state visible to all agents is defined as:
\begin{equation}
S_{app} = \{c_i, e_{i,j} \mid c_i, e_{i,j} \in G_A \}
\end{equation}
and the resource subgraph assigned to that local agent is:
\begin{equation}
S_{res}^{local} = \{v_i, l_{i,j} \mid v_i, l_{i,j} \in G_R^{local}\} 
\end{equation}








\textbf{Combined Observation Space for the Global Agent} \newline
The global agent maintains a high-level view of the entire system by aggregating local subgraph information: 

\begin{equation}
O_{global} = \{S_{app}, S_{res} = \bigcup_{i} aggr(S_{res}^{local}(i))\}
\end{equation}
where $S_{app}$ is the same overall application state as seen by the local agents and $S_{res}$ is the aggregated resource view composed of all local subgraphs. 


\subsection{Action Space}
The action space defines the possible decisions that local and global RL agents can make in the Cloud-Edge Continuum. It is designed to reflect the decentralized structure and enable scalability for dynamic environments.

\begin{itemize}
    \item \textbf{Local Agents:} Assign application components $c_i$ to resource nodes $v_j$ within their subgraph:
\begin{equation}
A_{local} = \{a_{ij} \mid a_{ij} \in \{0, 1\}, c_i \in C, v_j \in V_{local}\}
\end{equation}
Here $a_{ij} = 1$ indicates that $c_i$ is placed on $v_j$.
    \item \textbf{Global Agent:} Decides which local agents to activate for resource management tasks:
\begin{equation}
A_{global} = \{a_{i} \mid a_{i} \in \{0, 1\}, i = 1, \dots, N_{local}\}
\end{equation}
where $N_{local}$ is the number of local agents in the system.

\end{itemize}

The proposed strategy offers decentralized management, where local agents handle granular task placement within their neighborhoods while the global agent manages high-level coordination. 

\subsection{Optimization Techniques}
This section outlines potential optimization techniques that may be employed during the implementation phase to enhance resource management in the Cloud-Edge Continuum. \newline
\textbf{GNN Embeddings:} Methods like GraphSAGE compress node and edge information for both local and global agents. \newline
\textbf{Multi-Agent DRL:} Algorithms such as PPO and A3C coordinate decisions decisions across the Cloud-Edge Continuum, leveraging frameworks like Stable-Baselines3 \newline
\textbf{Experience Replay}: Stores experiences from pretraining and joint training for more stable and rapid convergence.

\subsection{Training Overview}
The training process consists of two phases: pretraining the local agents and jointly training the global agent alongside the local agents.

\textbf{Phase 1}: The GNN generates embeddings to create $O_{local_{i}}$, and local components are placed sequentially using local policies. The local agent policies are trained MARL.

\textbf{Phase 2}: The GNN generates embeddings by aggregating local information to represent the global state. The global agent then sequentially selects the local agent to execute the sequential placement strategy. The global and pretrained local agents are jointly trained to optimize a combined reward.

\section{Evaluation Plan}
\label{sec:pe}
To validate our approach, we plan to evaluate it in a simulated environment to analyze the system behavior in controlled settings before moving to real-world deployments.

\subsection{Simulation Environment}
A simulation framework like \textbf{ECLYPSE}\cite{Massa_ECLYPSE_an_Edge-Cloud_2024} could be used to verify the approach. Used for distributed computing systems, with an emphasis on edge-cloud infrastructure, it is a highly extensible framework and offers detailed application behavior modeling. We are considering extending it to create a tailored platform for implementing and testing MARL in distributed environments, enabling fine-grained control over simulation parameters.

\subsection{Evaluation Metrics}
The metrics previously defined in \ref{sec:cost_model}, RU, CT and SVR, will be used to assess the performance of the proposed system. Additionally, we will use \textbf{Training Efficiency}, defined as the computational efficiency of the training process including \textbf{Training Time} and \textbf{Memory Usage}, highlighting the feasibility of scaling to larger systems.


\section{Conclusion}
\label{sec:fw}
We presented our viewpoint on exploiting distributed local and global learning for scalable resource management in the Continuum platform, introducing a hierarchical system organization and an overview of the training process for the local and global agents. Building on the findings of the practical evaluation, we plan to validate our approach further through several steps. First, we aim to use real-world testbeds to validate the practicality and robustness of the proposed system. Second, we intend to develop hierarchical coordination and validate the system's scalability through simulations in larger-scale environments with more nodes, edges, and application components. Third, we plan to incorporate economic considerations into the optimization framework, comparing the cost-effectiveness of centralized and decentralized approaches.
%
    

To meet large-scale requirements, future work will explore \textbf{local agents collaboration} by integrating the action and state spaces of multiple agents, enabling more coordinated and efficient decision-making across the system.




\bibliographystyle{IEEEtran}
\bibliography{reference_abbrs}

\vspace{12pt}

\end{document}